\documentclass[reprint, superscriptaddress, amsmath, amssymb, aps, prb]{revtex4-2}
\usepackage[dvipdfmx]{graphicx}
\usepackage{color}
\usepackage{dcolumn}
\usepackage{bm}
\usepackage[dvipdfmx]{hyperref}
\usepackage{hyperref}
\usepackage{comment}
\usepackage{xcolor}
\usepackage{physics}

\begin{document}
\preprint{APS/123-QED}
\title{Robust realization of spin-polarized specular Andreev reflection \\ in V$_2$O-based altermagnets}
\author{Yutaro Nagae}
\affiliation{Department of Applied Physics, Nagoya University, Nagoya 464-8603, Japan}

\author{Andreas P. Schnyder}
\affiliation{Max-Planck-Institut f\"{u}r Festk\"{o}rperforschung, Heisenbergstra\ss e 1, D-70569 Stuttgart, Germany}

\author{Satoshi Ikegaya}
\affiliation{Department of Applied Physics, Hokkaido University, Sapporo 060-8628, Japan}

\date{\today}

\begin{abstract}
We theoretically investigate charge transport in a junction between a conventional superconductor
and a V$_2$O-based altermagnet exhibiting distinctive spin-split quasi-one-dimensional Fermi surfaces.
The altermagnet is described by a microscopically motivated seven-basis-state model with six-orbital characters
that incorporates sublattice degrees of freedom associated with both V and O sites.
Based on calculations performed under various boundary conditions, we demonstrate the robust emergence of specular Andreev reflection with a distinctive spin polarization.
Furthermore, we propose an efficient multiterminal setup to detect this specular Andreev reflection through nonlocal conductance measurements.
Our results establish V$_2$O-based altermagnets as a promising platform for realizing spin-resolved Cooper pair splitting, which is essential for generating energy-entangled electron pairs.
\end{abstract}
\maketitle

\section{Introduction}
Andreev reflection (AR) is a fundamental scattering process at normal-metal--superconductor interfaces,
in which an incident electron is reflected as a hole, while a Cooper pair is injected into the superconductor~\cite{Andreev_1964}.
Conventional AR is retroreflective, meaning that the reflected hole retraces the trajectory of the incoming electron.
This is in contrast to nonretroreflective AR processes, namely crossed AR (CAR)~\cite{Deutscher_2000,Recher_2001} and specular AR (SAR)~\cite{Beenakker_2006}.
In CAR, an electron from one lead is converted into a hole in a spatially separated lead,
whereas in SAR, the electron is specularly reflected, due to the unconventional band structure of the normal segment.
The reverse processes of these nonretroreflective ARs correspond to Cooper-pair splitting,
which is of great interest as a source of nonlocal entangled electron pairs for quantum technologies~\cite{Lesovik_2001, Martin_2002}.
While CAR has been experimentally demonstrated in various semiconductor--superconductor hybrid systems~\cite{Hofstetter_2009,Hofstetter_2011,Schindele_2012,Das_2012,Fulop_2014,Fulop_2015,Baba_2018,Wang_2022,Bordoloi_2022,Jong_2023}, SAR remains largely unexplored.
This is likely because SAR has thus far been predicted only in a limited class of systems, such as Dirac or Weyl semimetals with finely tuned chemical potentials%
~\cite{Beenakker_2006,chen_13,asgari_16,sun_17,sun_17,sun_20}.

To overcome this limitation,
we theoretically demonstrated in our previous work~\cite{Nagae_2025} that SAR can occur at junctions between a conventional superconductor and an altermagnet,
a novel class of magnetic materials~\cite{Naka_2019,Hayami_2019,Hayami_2020,Naka_2020,Sinova_2020,Naka_2021,Mazin_2021,Seo_2021,Smejkal_2022,Smejkal_2022_2}.
Remarkably, SAR in an altermagnet is accompanied by a distinctive spin polarization,
giving rise to a novel functionality that combines a Cooper-pair splitter with a spin beam splitter.
However, in that study, the altermagnet was modeled within a minimal single-orbital framework,
in which the anisotropic spin-split band structure originates merely from an anisotropic exchange potential.
Thus, the intrinsic collinear magnetic order%
---characterized by spin-point-group symmetry and representing the true origin of the anisotropic spin-split bands~\cite{Hayami_2019,Smejkal_2022}---%
was not explicitly included.
Although transport in altermagnet--superconductor junctions has been widely discussed~\cite{Linder_2023,Beenakker_2023,Papaj_2023,Giil_2024,Kazmin_2025,Fukaya_2025,Fukaya_2025_2},
calculations that explicitly incorporate the magnetic order remain extremely limited.
Meanwhile, it has been suggested that transport in antiferromagnet--superconductor junctions can be highly sensitive to the magnetic configuration at the interface%
~\cite{bobkova_2005,barach_2005,barach_2006,kamra_2022}.
In this context, developing a transport theory for altermagnet--superconductor junctions
that explicitly accounts for the intrinsic magnetic order is essential to establish a firm theoretical foundation for future experimental investigations in this emerging field.

In this paper, we focus on V$_2$O-based metallic altermagnets,
including KV$_2$Se$_2$O, Rb$_{1-\delta}$V$_2$Te$_2$O, Cs$_{1-\delta}$V$_2$Te$_2$O, and CsV$_2$Se$_2$O, as promising candidate materials~\cite{Jiang_2025,Zhang_2025,Lai_2025,Liu_2025,WZhang_2025,ZWang_2025,Fu_2025,QHu_2026,XCheng_2026}.
In these compounds, distinctive spin-split quasi-one-dimensional Fermi surfaces, originating primarily from $d$-orbital electrons in the V$_2$O planes, have been reported.
Here we employ a microscopically motivated seven-basis-state model with six-orbital characters that explicitly describes the V$_2$O plane,
incorporating not only the magnetic vanadium sites hosting $d$-orbital electrons, but also the oxygen sites with $p$-orbital electrons.
By performing extensive calculations of scattering coefficients under various boundary conditions,
we demonstrate that spin-polarized SAR emerges at the junction between the V$_2$O-based altermagnet and a conventional superconductor.
In addition, we propose a multiterminal setup to experimentally detect SAR via nonlocal conductance measurements.
Importantly, by tuning the bias-voltage configuration,
our proposed setup enables access to both regimes in which SAR either contributes or does not contribute to nonlocal charge currents,
thereby allowing direct comparison between the SAR signature and a trivial control signal.
Consequently, our results demonstrate that V$_2$O-based altermagnets constitute a promising platform for realizing spin-resolved Cooper-pair splitting,
representing an important step toward the creation of energy-entangled electron pairs~\cite{Lesovik_2001}.

\section{Model and Formulation}

\subsection{V$_2$O-based Altermagnet}
\begin{figure}[tttt]
\begin{center}
\includegraphics[width=0.5\textwidth]{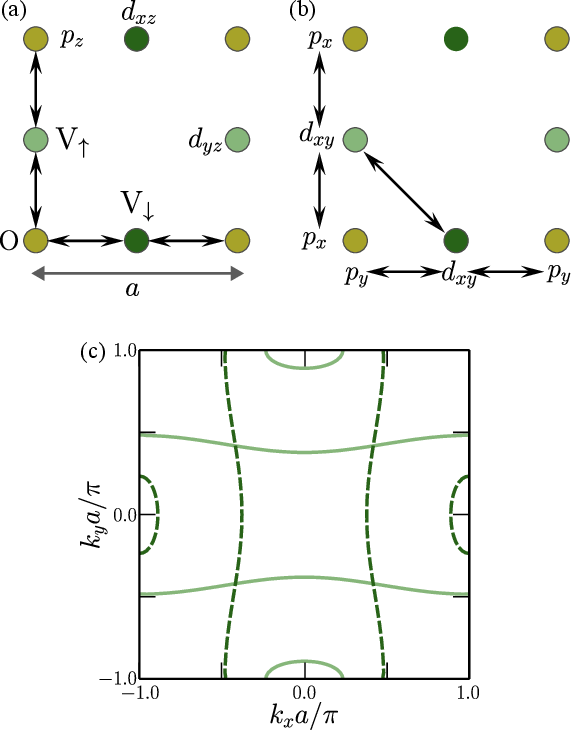}
\caption{(a), (b) Schematic illustration of the hopping integrals in the V$_2$O plane described by $H_{\textrm{AM},1}$ and $H_{\textrm{AM},2}$, respectively.
(c) FSs obtained by diagonalizing the full Hamiltonian $H_{\textrm{AM}}$.
The solid (dashed) lines indicate the spin-up (spin-down) FSs.}
\label{figure1}
\end{center}
\end{figure}

First, we introduce the model employed to describe the V$_2$O-based altermagnet.
The structure of the V$_2$O plane, which forms a Lieb lattice, is summarized in Figs.~\ref{figure1}(a)~and~~\ref{figure1}(b).
Nearest-neighbor vanadium (V) sites carry opposite spins aligned along the $c$ axis.
In the following, we denote these magnetic vanadium sites as V$_{\uparrow}$ and V$_{\downarrow}$.
The edge-centered oxygen (O) sites break the rotational symmetry of the magnetic V sublattice,
while the magnetic structure preserves a combined spin--lattice $C_{4z}\mathcal{T}$ symmetry, characteristic of $d$-wave altermagnetism~\cite{Brekke_2023}.
For example, angle-resolved photoemission spectroscopy measurements on KV$_2$Se$_2$O report four distinctive Fermi surfaces (FSs)~\cite{Jiang_2025}, as shown in Fig.~\ref{figure1}(c):
a spin-down (spin-up) quasi-one-dimensional FS extending along $k_y$ ($k_x$) direction,
primarily formed by $d_{xz}$ orbital ($d_{yz}$ orbital) electrons at V$_{\downarrow}$ (V$_{\uparrow}$) sites;
and a spin-down (spin-up) quasi-two-dimensional FS surrounding the Y (X) point at the Brillouin-zone boundary,
primarily formed by $d_{xy}$ orbital electrons at V$_{\downarrow}$ (V$_{\uparrow}$) sites.
To reproduce these Fermi surfaces, we employ a two-dimensional seven-basis tight-binding model
that includes the $d_{xz}$ orbital at the V$_{\downarrow}$ sites,
the $d_{yz}$ orbital at the V$_{\uparrow}$ sites,
the $d_{xy}$ orbitals at both V$_{\uparrow}$ and V$_{\downarrow}$ sites
(denoted as $d_{xy}\textrm{V}_{\uparrow}$ and $d_{xy}\textrm{V}_{\downarrow}$ orbitals, respectively),
as well as the $p_x$, $p_y$, and $p_z$ orbitals at the O sites.
Although the oxygen $p$ orbitals contribute less to the FSs, 
we explicitly describe these O sites because they can affect the boundary conditions at the interface.
Considering a minimal set of hopping integrals up to essential second-nearest neighbors,
the Hamiltonian in momentum space is given by:
\begin{align}
H_\textrm{AM} = H_{\textrm{AM},1} + H_{\textrm{AM},2},
\label{eq1}
\end{align}
with
\begin{align}
\begin{split}
&H_{\textrm{AM},1} = \sum_{\vb*{k}}\sum_\sigma \vb*{\Psi}_{\vb*{k},\sigma,1}^\dagger \mathcal{H}_{\vb*{k},\sigma,1}\vb*{\Psi}_{\vb*{k},\sigma,1}, \\
&\vb*{\Psi}_{\vb*{k},\sigma,1} = \qty[c_{\vb*{k},\sigma,p_z}, c_{\vb*{k},\sigma,d_{yz}}, c_{\vb*{k},\sigma,d_{xz}}]^{\mathrm{T}},\\
&\mathcal{H}_{\vb*{k},\sigma,1} = 
\mqty(-\varepsilon_1-\mu & h_{p_z\text{-}d_{yz}} & h_{p_z\text{-}d_{xz}}
\\ h^{\ast}_{p_z\text{-}d_{yz}} & -s_\sigma m_1-\mu & 0 \\ h^{\ast}_{p_z\text{-}d_{xz}} & 0 & s_\sigma m_1 - \mu ),\\
&h_{p_z\text{-}d_{yz}}= 2i t_1 e^{-\frac{ik_ya}{2}} \sin \frac{k_ya}{2}, \\
&h_{p_z\text{-}d_{xz}}= 2i t_1 e^{-\frac{ik_xa}{2}}\sin \frac{k_xa}{2},
\end{split}
\end{align}
and
\begin{align}
&H_{\textrm{AM},2} = \sum_{\vb*{k}}\sum_\sigma \vb*{\Psi}_{\vb*{k},\sigma,2}^\dagger \mathcal{H}_{\vb*{k},\sigma,2}\vb*{\Psi}_{\vb*{k},\sigma,2}, \nonumber\\
&\vb*{\Psi}_{\vb*{k},\sigma,2} =
\qty[c_{\vb*{k},\sigma,p_x}, c_{\vb*{k},\sigma,p_y},
c_{\vb*{k},\sigma,d_{xy}\textrm{V}_{\uparrow}},
c_{\vb*{k},\sigma,d_{xy}\textrm{V}_{\downarrow}}]^{\mathrm{T}}, \nonumber\\
&\mathcal{H}_{\vb*{k},\sigma,2} =
\mqty(-\varepsilon_2-\mu & 0 & 0 & h_{p_x\text{-}d_{xy}}\\
0 & -\varepsilon_2 - \mu & h_{p_y\text{-}d_{xy}} & 0 \\
0 & h_{p_y\text{-}d_{xy}}^{\ast} & -s_\sigma m_2 -\mu & h_{d_{xy}\text{-}d_{xy}} \\
h_{p_x\text{-}d_{xy}}^{\ast} & 0 & h_{d_{xy}\text{-}d_{xy}}^{\ast} & s_\sigma m_2-\mu),\nonumber\\
&h_{p_y\text{-}d_{xy}}= 2i t_2 e^{-\frac{ik_ya}{2}} \sin \frac{k_ya}{2}, \nonumber\\
&h_{p_x\text{-}d_{xy}} = 2i t_2 e^{-\frac{ik_xa}{2}} \sin \frac{k_xa}{2},\nonumber\\
&h_{d_{xy}\text{-}d_{xy}} = -4 t_3 e^{-\frac{i(k_x-k_y)a}{2}} \cos \frac{k_xa}{2} \cos \frac{k_ya}{2},
\end{align}
where the Hamiltonian can be divided into two subspaces describing, respectively,
the quasi-one-dimensional FSs extending along the $k_x$ and $k_y$ directions (i.e., $H_{\textrm{AM},1}$)
and the quasi-two-dimensional FSs located at the Brillouin-zone boundary (i.e., $H_{\textrm{AM},2}$).
Here $c_{\vb*{k}, \sigma, \alpha}^\dagger(c_{\vb*{k}, \sigma, \alpha})$ represents the creation (annihilation) operator 
of an electron with momentum $\vb*{k}$, spin $\sigma$, and orbital $\alpha$.
$\mu$ is the chemical potential,
$\varepsilon_{1(2)}$ denotes the on-site potential for the $p_z$ orbital ($p_x$ and $p_y$ orbitals) at the O sites,
and $m_{1(2)}$ represents the exchange potential at the V$_\sigma$ sites for the $d_{xz}$ and $d_{yz}$ orbitals ($d_{xy}$ orbital).
$t_1$ is the nearest-neighbor hopping integral between
the $p_z$-orbital at the O site and the $d_{xz}$ orbital at the V$_{\uparrow}$ site or the $d_{yz}$-orbital at the V$_{\downarrow}$ site,
$t_2$ is the nearest-neighbor hopping integral between
the $p_{x}$ or $p_{y}$ orbitals at the O sites and $d_{xy}$ orbital at the V$_\sigma$ sites,
$t_3$ is the next-nearest-neighbor hopping integral between $d_{xy}$ orbitals at the V$_\sigma$ sites,
and $s_{\uparrow(\downarrow)} = +1(-1)$.
The lattice constant is denoted by $a$.
In the following, we fix the parameters as $\mu = -0.8t_1, \varepsilon_1 = 5t_1, m_1 = 1.1t_1, t_2 = 0.5t_1, \varepsilon_2 = -0.5t_1, m_2 = 0.7t_1$, and $t_3 =0.4t_1$,
which are used to generate Fig.~\ref{figure1}(c).
Note that the FSs obtained from $H_{\textrm{AM}}$ well reproduce those observed in experiments and first-principles calculations of KV$_2$Se$_2$O~\cite{Jiang_2025}.

\subsection{Multiterminal setup}
\begin{figure}[tttt]
\begin{center}
\includegraphics[width=0.5\textwidth]{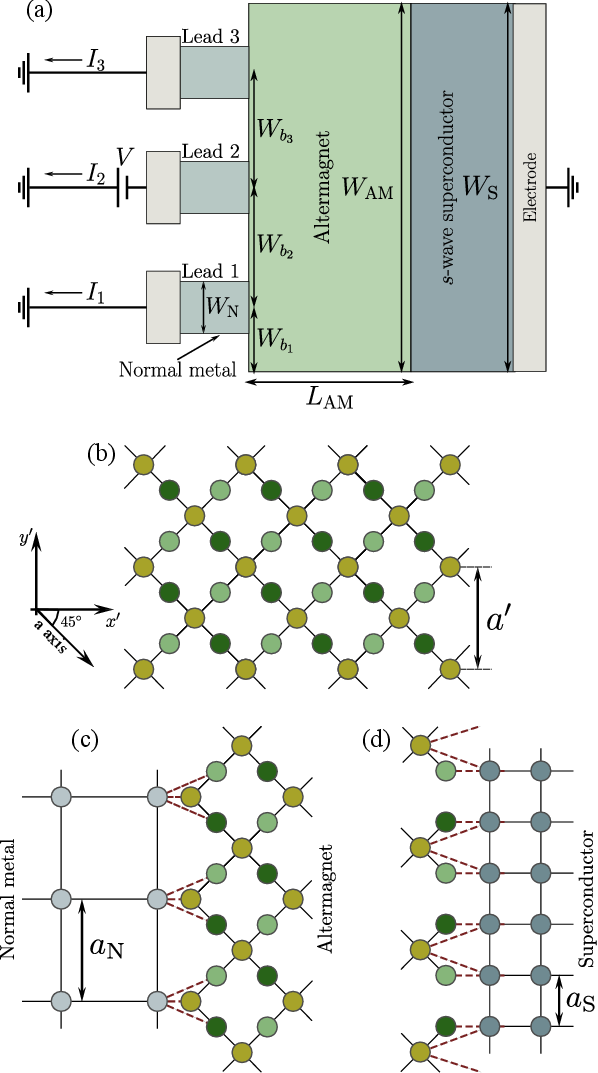}
\caption{(a) Schematic illustration of the multiterminal setup consisting of three normal-metal leads, the V$_2$O-based altermagnet, and an $s$-wave superconductor.
(b) Schematic illustration of the altermagnet tilted by $45^\circ$.
(c), (d) Lattice structures at the interfaces between the normal metal and the altermagnet, and between the altermagnet and the superconductor, respectively.}
\label{figure2}
\end{center}
\end{figure}

In this paper, we study the transport properties of a multiterminal setup consisting of
three normal metal leads, the V$_2$O-based altermagnet, and a conventional spin-singlet $s$-wave superconductor, shown in Fig.~\ref{figure2}(a).
To realize SAR at the altermagnet--superconductor interface, following our previous study~\cite{Nagae_2025},
the altermagnet is tilted by $45^\circ$ with respect to the $a$ axis of the V$_2$O plane, as shown in Fig.~\ref{figure2}(b).
As discussed later, the normal-metal leads are attached to detect SAR in this setup.
As illustrated in Fig.~\ref{figure2}(a), the first normal-metal lead is positioned at a distance $W_{b_1}$ from the bottom edge of the altermagnet.
The second and third leads are attached at intervals of $W_{b_2}$ and $W_{b_3}$, respectively.
The width of the normal metal leads, the altermagnet, and the superconductor are denoted by $W_\textrm{N}$, $W_\textrm{AM}$, and $W_\textrm{S}$, respectively.
The length of the altermagnet segment is $L_\textrm{AM}$.
As shown in Figs.~\ref{figure2}(c)~and~\ref{figure2}(d),
we assume that the normal metals and the superconductor are described by square lattice tight-binding models
with lattice constants $a_{\textrm{N}}=a^{\prime}$ and $a_{\textrm{S}}=a^{\prime}/2$ with $a^{\prime}=\sqrt{2}a$, respectively.
Note that, as demonstrated later, the qualitative results are insensitive to the details of the junction interfaces.
\begin{figure*}[tttt]
\begin{center}
\includegraphics[width=\textwidth]{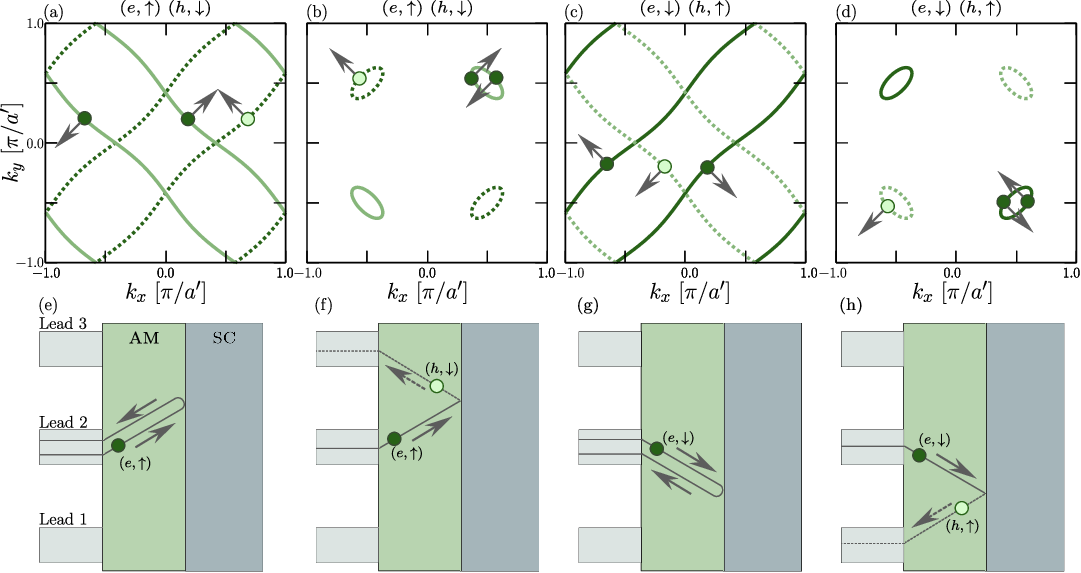}
\caption{(a)–(d) FSs of the altermagnet tilted by $45^\circ$.
(a) [(b)] show the quasi-one-dimensional [quasi-two-dimensional] FSs of spin-up electrons and spin-down holes,
while (c) [(d)] show those of spin-down electrons and spin-up holes.
The solid and dashed lines represent the FSs of electrons and holes, respectively,
and the arrows indicate the directions of the corresponding group velocities.
(e)–(h) Schematic illustrations of the possible scattering processes in the present device.
In (e) and (f) [(g) and (h)], spin-up [spin-down] electrons are incident from the second normal lead.}
\label{figure3}
\end{center}
\end{figure*}
We describe the present setup by the two-dimensional tight-binding Hamiltonian $H=H_{\textrm{AM,tilted}}+H_\textrm{N}+H_\textrm{S}+H_\textrm{AM-N}+H_\textrm{S-AM}$,
whose explicit form is given in Appendix~\ref{sec:appendixa}.
In the calculations below, the nearest-neighbor hopping integrals (chemical potentials) in the normal and superconducting segments
are fixed to $t_{\textrm{N}} = t_{\textrm{S}} = t_1$ ($\mu_{\textrm{N}} = \mu_{\textrm{S}} = -t_1$),
and the spin-singlet $s$-wave pair potential in the superconducting segment is fixed to $\Delta = 0.001t_1$.
The terms $H_\textrm{N-AM}$ and $H_{\textrm{AM-S}}$ describe the interfaces between the normal metals and the altermanget,
and between the altermagnet and the superconductor, respectively.
We assume the hopping terms in $H_\textrm{N-AM}$ ($H_{\textrm{AM-S}}$) connect
the outermost sites of the normal metal (superconductor) to the outermost O sites and magnetic V sites of the altermagnet,
as indicated by the dashed lines in Figs.~\ref{figure2}(c)~and~\ref{figure2}(d), respectively.
In practical experimental conditions, some degree of surface roughness, crystal misorientation, and lattice mismatch is inevitable.
To effectively capture such effects, we assume the hopping integrals between the normal metal (superconductor) and the altermagnet,
$t_{J_{\textrm{N}(\textrm{S})},\alpha ;\vb*{r}, \vb*{r}'}$, are randomly distributed as
\begin{align}
t_{J_\textrm{N(S)},\alpha;\vb*{r},\vb*{r}'} = \bar{t}_{J_\textrm{N(S)},\alpha} + \delta t_{J_\textrm{N(S)},\alpha;\vb*{r},\vb*{r}'},
\end{align}
for the allowed hoppings.
Here, $\bar{t}_{J_\textrm{N(S)},\alpha}$ denotes the mean value of the hopping amplitude to the $\alpha$ orbital,
while $\delta t_{J_\textrm{N(S)},\alpha;\vb*{r},\vb*{r}'}$ is uniformly distributed in the interval $[-{\delta t_{J_\textrm{N(S)},\alpha}}/2,\; {\delta t_{J_\textrm{N(S)},\alpha}}/2]$.
For more details regarding the Hamiltonian, see Appendix~\ref{sec:appendixa}.

\subsection{Formulation} 
In this paper, we discuss the differential conductance in the present setup. 
The bias voltage $V$ is applied to the second normal lead, while the other normal leads and the superconductor are grounded.
The differential conductance in the $\alpha$-th lead is defined as $G_{\alpha} = dI_\alpha/dV$, where $I_\alpha$ denotes the charge current flowing in the $\alpha$-th normal metal lead.
Within the Blonder-Tinkham-Klapwijk (BTK) formalism, the differential conductance at zero temperature is calculated as~\cite{Blonder_1982,Anantram_1996},
\begin{align}
\begin{split}
& G_{\alpha}(eV) = G_{\alpha,\uparrow}(eV) + G_{\alpha,\downarrow}(eV),\\
& G_{\alpha,\sigma}(eV) = \frac{e^2}{h}\mathrm{Tr}\qty[\delta_{\alpha,2}\hat{\vb*{1}} - \hat{R}_{\alpha 2,\sigma}^e + \hat{R}_{\alpha 2,\sigma}^h]_{E=eV}, \\
& \hat{R}_{\alpha\beta,\sigma}^\nu = \sum_{\sigma'} \hat{s}_{\alpha\beta,\sigma\sigma'}^{\nu e}\qty(\hat{s}_{\alpha\beta,\sigma\sigma'}^{\nu e})^\dagger, \ (\nu = e, h),
\end{split}
\end{align}
where $\hat{\vb*{1}}$ is the $N_c\times N_c$ identity matrix with $N_c$ denoting the number of propagating channels per spin in the normal metal lead,
and $\hat{s}_{\alpha\beta,\sigma\sigma'}^{ee}$ ($\hat{s}_{\alpha\beta,\sigma\sigma'}^{he}$) is an $N_c\times N_c$ matrix containing the scattering coefficients
from an electron with spin $\sigma'$ in the $\beta$-th normal metal lead to an electron (hole) with spin $\sigma$ in the $\alpha$-th normal metal lead at energy $E$.
The scattering coefficients are calculated using the recursive Green's function technique~\cite{Lee_1981,Ando_1991}.
Note that the BTK formalism is quantitatively justified for bias voltages well below the superconducting gap,
where the effect of voltage drop inside the altermagnet segment is negligible.
We emphasize that the nonlocal conductances $G_1$ and $G_3$ become positive only when interlead Andreev scattering,
mainly originating from SAR at the altermagnet--superconductor interface, dominates the transport between the leads.

\section{Results}
\subsection{Positive nonlocal conductance} \label{sec:3a}
In this section, we investigate the transport properties when the normal leads are symmetrically positioned, as illustrated in the lower panels of Fig.~\ref{figure3},
where $W_{b_2}=W_{b_3}$ and $2(W_{b_1}+W_{b_2})=W_{\textrm{AM}}$.
In the upper panels of Fig.~\ref{figure3}, we show the FSs of the altermagnet tilted by $45^\circ$.
The solid and dashed lines represent the FSs of electrons and holes, respectively.
The arrows indicate the directions of the corresponding group velocity.
The $C_{4z}\mathcal{T}$ symmetry of the altermagnet,
together with the intrinsic particle-hole symmetry of the superconducting system,
ensures that the FSs of spin-up (spin-down) electrons overlap with those of spin-down (spin-up) holes after a relative $90^\circ$ rotation.
Figures~\ref{figure3}(a) and \ref{figure3}(b) show the FSs for spin-up electrons and for spin-down holes.
For both quasi-one-dimensional FSs in Fig.~\ref{figure3}(a) and the quasi-two-dimensional FSs in Fig.~\ref{figure3}(b),
right-moving spin-up electrons tend to have group velocities along the $(x+y)$ direction,
whereas left-moving spin-up electrons tend to have group velocities along the $-(x+y)$ direction.
Meanwhile, left-moving spin-down holes tend to have group velocities along the $-(x-y)$ direction.
Given this configuration, as illustrated in Fig.~\ref{figure3}(e), the normal reflection at the altermagnet--superconductor interface becomes retroreflective.
In contrast, remarkably, SAR occurs at the altermagnet--superconductor interface, giving rise to inter-lead Andreev scattering from the second lead to the third lead.
Similarly, for a spin-down incident electron, SAR leads to inter-lead Andreev scattering from the second lead to the first lead, while the normal reflection remains retroreflective,
as illustrated in Figs.~\ref{figure3}(g) and 3(h), respectively.
Consequently, the charge current from the second to the third (first) lead is governed by spin-polarized SAR,
resulting in a positive nonlocal conductance, $G_{3} \sim G_{3,\downarrow}>0$ ($G_{1} \sim G_{1,\uparrow}>0$).
In the present model, the $45^\circ$ interface orientation is chosen
because it is expected to maximize the interlead scattering mediated by SAR between the symmetrically positioned leads, as illustrated by the scattering processes in Fig.~\ref{figure3}.
Importantly, this behavior follows solely from the characteristic structure of the spin-split Fermi surfaces of the altermagnet
and does not depend on the microscopic details of the junction interface.
Thus, the realization of positive nonlocal conductance is expected to be insensitive to the specific boundary conditions of the junction.
\begin{figure}[tttt]
\begin{center}
\includegraphics[width=0.5\textwidth]{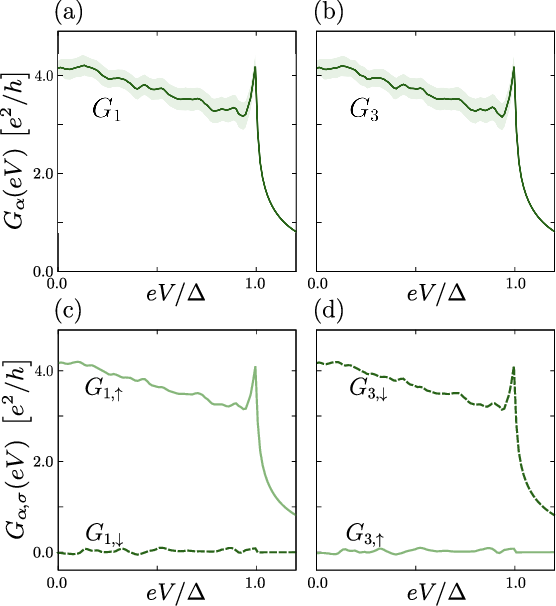}
\caption{(a), (b) Nonlocal differential conductances $G_1$ and $G_3$ as functions of the bias voltage $eV$.
The spin-resolved nonlocal conductances $G_{1,\sigma}$ and $G_{3,\sigma}$ are shown in (c) and (d), respectively.
The solid lines and shaded regions in (a) and (b) represent the averages and the standard deviations, respectively, obtained from 1000 realizations of the surface roughness.}
\label{figure4}
\end{center}
\end{figure}
\begin{figure*}[tttt]
\begin{center}
\includegraphics[width=0.75\textwidth]{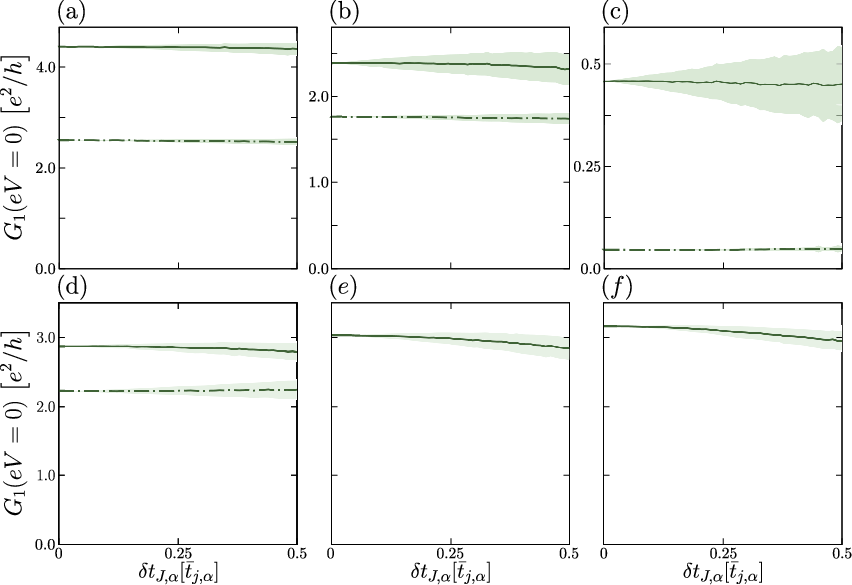}
\caption{Nonlocal differential conductance $G_1$ at zero bias voltage as a function of the randomness strength in the hopping amplitude at the junction (i.e. $\delta t_{J_{\textrm{N(S)},\alpha}}$) under various interfacial conditions.
In (a), the solid line corresponds to $\bar{t}_{J_\textrm{N(S)},\alpha}=0.5t_1$ and the dash-dotted line corresponds to $\bar{t}_{J_\textrm{N(S)},\alpha}=0.25t_1$ for $\alpha = p_z, d_{xz}, d_{yz}$, respectively, while $\bar{t}_{J_\textrm{N(S)},\alpha}={\delta t_{J_\textrm{N(S)},\alpha}}=0$ for all other orbitals.
In (b), the solid line corresponds to $\bar{t}_{J_\textrm{N(S)},\alpha}=0.5t_1$ and the dash-dotted line corresponds to $\bar{t}_{J_\textrm{N(S)},\alpha}=0.25t_1$ for $\alpha = d_{xz},d_{yz},d_{xy}$V$_\uparrow,d_{xy}$V$_\downarrow$, respectively, while $\bar{t}_{J_\textrm{N(S)},\alpha}={\delta t_{J_\textrm{N(S)},\alpha}}=0$ for all other orbitals.
In (c), the solid line corresponds to $\bar{t}_{J_\textrm{N(S)},\alpha}=0.5t_1$ and the dash-dotted line corresponds to $\bar{t}_{J_\textrm{N(S)},\alpha}=0.25t_1$ for $\alpha = p_x, p_y, p_z$, respectively, while $\bar{t}_{J_\textrm{N(S)},\alpha}={\delta t_{J_\textrm{N(S)},\alpha}}=0$ for all other orbitals.
In (d), the solid line corresponds to $\bar{t}_{J_\textrm{N(S)},\alpha}=0.5t_1 (0.25t_1)$ and the dash-dotted line corresponds to $\bar{t}_{J_\textrm{N(S)},\alpha}=0.25t_1 (0.5t_1)$, respectively.
In (e), we set $\bar{t}_{J_\textrm{N(S)},\alpha}=0.25t_1$ for the first lead---altermagnet interface, and $\bar{t}_{J_\textrm{N(S)},\alpha}=0.5t_1$ for the other lead---altermagnet interfaces.
In (f), we choose $\bar{t}_{J_\textrm{N(S)},\alpha}=0.5t_1$ for spin-up electrons, and $\bar{t}_{J_\textrm{N(S)},\alpha}=0.25t_1$ for spin-down electrons.
The solid lines and the dash-dotted lines represent the averages, 
and shaded regions represent the standard deviations.}\label{figure5}
\end{center}
\end{figure*}
\begin{figure}[tttt]
\begin{center}
\includegraphics[width=0.5\textwidth]{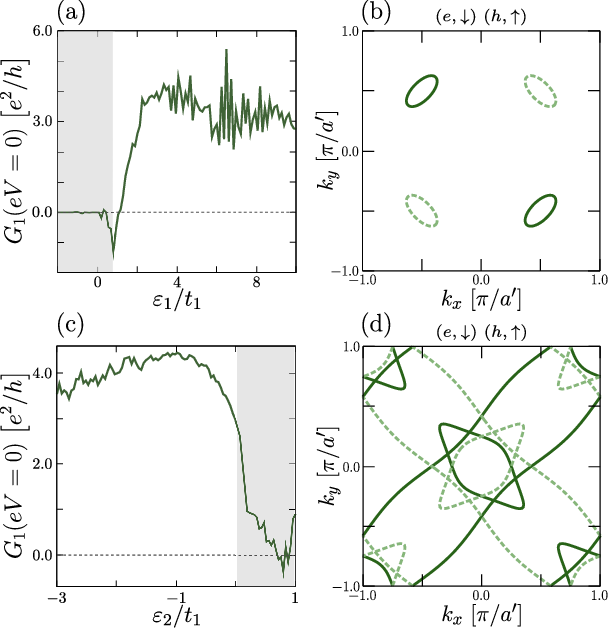}
\caption{(a) $G_1$ as a function of the on-site energy for the $p_z$ orbital at the O sites (i.e. $\varepsilon_1$).
In the shaded region satisfying $\varepsilon_1<-\mu$, the altermagnet does not exhibit quasi-one-dimensional FSs.
(b) Fermi surfaces with the on-site energy for $p_z$ orbital at O sites as $\varepsilon_1=-2t_1 <-\mu$.
(c) $G_1$ as a function of the on-site energy for the $p_x$ and $p_y$ orbitals at the O sites (i.e. $\varepsilon_2$).
(d) Fermi surfaces with the on-site energy for $p_x$ and $p_y$ orbital at O sites as $\varepsilon_2=t_1>-\mu-4t_2^2/(m_2-\mu)$.
}
\label{figure6}
\end{center}
\end{figure}

Figure~\ref{figure4} shows the numerical results for the nonlocal differential conductance as a function of the bias voltage.
In Fig.~\ref{figure4}(a) [\ref{figure4}(b)], we present $G_1$ [$G_3$],
while the spin-resolved conductances $G_{1,\sigma}$ [$G_{3,\sigma}$] are shown in Fig.~\ref{figure4}(c) [\ref{figure4}(d)].
The parameters are chosen as $W_{\textrm{AM}}=200a',L_\textrm{AM} = 40a', W_\textrm{N}=20a',W_\textrm{S}=199a', W_{b_1}=(41/\sqrt{2})a', W_{b_2} = W_{b_3} = 80a'$.
The hopping integrals at the junction interfaces are assumed to be independent of the orbital degrees of freedom,
with $\bar{t}_{J_\textrm{N(S)},\alpha}=0.5t_1$ and $\delta t_{J_\textrm{N(S)},\alpha}=0.5\bar{t}_{J_\textrm{N(S)},\alpha}$.
In Figs.~\ref{figure4}(a) and \ref{figure4}(b), the solid curves represent the averages, and the shaded regions show the associated standard deviations,
obtained from $10^3$ independent realizations of the interface hopping integral, ${\delta t_{J_\textrm{N(S)},\alpha;\vb*{r},\vb*{r}'}}$.
In accordance with our expectation, we clearly observe significant positive nonlocal conductances, i.e., $G_{1} \sim G_{1,\uparrow}$ and $G_{3} \sim G_{3,\downarrow}$.
We emphasize that our previous work~\cite{Nagae_2025} already demonstrated positive nonlocal conductance in a single-orbital model of altermagnets
that excludes both the O orbitals and the $d_{xy}$ orbital.
Therefore, the results shown in Fig.~\ref{figure4} demonstrate that SAR in altermagnets remains robust even in a more realistic model
that explicitly incorporates the V and O lattice sites of the V$_2$O plane and reproduces the Fermi surfaces of V$_2$O-based altermagnets,
consistent with previous experimental observations and first-principles calculations~\cite{Jiang_2025}.
As shown in Fig.~\ref{figure4}, substantial positive nonlocal conductance is expected throughout the subgap regime,
whereas the BTK formalism is quantitatively reliable only at bias voltages well below the superconducting gap.
A self-consistent nonequilibrium treatment applicable beyond this regime remains an important subject for future work.

Figure~\ref{figure5} shows the zero-bias differential conductance $G_1$ as a function of the randomness strength in the hopping amplitude at the junction $\delta t_{J_\textrm{N(S)},\alpha}$.
In Fig.~\ref{figure5}(a), we assume that the normal-metal and superconducting leads couple only to 
the $d_{xz}$ orbital at the V$_{\downarrow}$ sites, the $d_{yz}$ orbital at the V$_{\uparrow}$ sites, and the $p_z$ orbital at the O sites.
In this case, transport in the altermagnet segment is governed by the quasi-one-dimensional FSs.
Specifically, for $\alpha = p_z, d_{xz}, d_{yz}$, 
we set $\bar{t}_{J_\textrm{N(S)},\alpha}=0.5t_1 (0.25t_1)$ for the solid (dash-dotted) line, 
while $\bar{t}_{J_\textrm{N(S)},\alpha}={\delta t_{J_\textrm{N(S)},\alpha}}=0$ for all other orbitals.
We confirm that the positive nonlocal conductance remains even in this case.
Note that the orbitals forming the quasi-one-dimensional FSs extend substantially along the out-of-plane direction.
Accordingly, although our analysis is restricted to two-dimensional planar junctions for numerical tractability,
the positive nonlocal conductance may also be realized when the normal-metal and superconducting leads are attached to the top surface of the V$_2$O-based altermagnet.
In Figs.~\ref{figure5}(b) and (c), 
we examine the nonlocal conductance for orbital-dependent interfacial couplings.
In Fig.~\ref{figure5}(b), the normal-metal and superconducting leads couple only to the magnetic V sites;
for $\alpha = d_{xz},d_{yz},d_{xy}$V$_\uparrow,d_{xy}$V$_\downarrow$,
we choose $\bar{t}_{J_\textrm{N(S)},\alpha}=0.5t_1 (0.25t_1)$ for the solid (dash-dotted) line,
while $\bar{t}_{J_\textrm{N(S)},\alpha}={\delta t_{J_\textrm{N(S)},\alpha}}=0$ for the other orbitals.
In Fig.~\ref{figure5}(c), the normal-metal and superconducting leads couple only to the O sites;
for $\alpha = p_x, p_y, p_z$, we set $\bar{t}_{J_\textrm{N(S)},\alpha}=0.5t_1$($0.25t_1$) for the solid (dash-dotted) line,  
while $\bar{t}_{J_\textrm{N(S)},\alpha}={\delta t_{J_\textrm{N(S)},\alpha}}=0$ for the other orbitals.
These results show that the positive nonlocal conductance is insensitive to the orbital dependence of the interfacial hopping integrals.
Next, we consider the differential conductance in the presence of contact asymmetries.
In Fig.~\ref{figure5}(d),  
we consider a significant imbalance in the hopping integrals between the normal metal--altermagnet and altermagnet--superconductor interfaces;
we set $\bar{t}_{J_\textrm{N(S)},\alpha} = 0.5t_1 (0.25t_1)$ for the solid line, 
and $\bar{t}_{J_\textrm{N(S)},\alpha} = 0.25t_1 (0.5t_1)$ for the dash-dotted line, independent of the orbital degrees of freedom.
In Fig.~\ref{figure5}(e), 
we introduce a contact asymmetry at the normal metal--altermagnet interfaces; 
we set $\bar{t}_{J_\textrm{N(S)},\alpha} = 0.25t_1$ for the first lead, 
while $\bar{t}_{J_\textrm{N(S)},\alpha} = 0.5t_1$ for the other leads.
We find that the positive nonlocal conductance remains even in the presence of contact asymmetries.
In Fig.~\ref{figure5}(f), 
we further introduce spin-dependent hopping integrals independent of the orbital degrees of freedom, 
chosen as $\bar{t}_{J_\textrm{N(S)},\alpha}=0.5t_1 (0.25t_1)$ for spin-up (spin-down) electrons.
The positive nonlocal conductance remains under spin-dependent interfacial conditions.
Taken together, these results demonstrate the robust realization of the positive nonlocal conductance under various interfacial conditions.

Figure~\ref{figure6}(a) shows the zero-bias differential conductance $G_1$ as a function of the on-site energy for the $p_z$ orbital ($\varepsilon_1$).
In the shaded region satisfying $\varepsilon_1<-\mu$, 
the quasi-one-dimensional FSs disappear, as illustrated for $\varepsilon_1=-2t_1$ in Fig.~\ref{figure6}(b).
In this region, as shown in Fig.~\ref{figure6}(a),
the nonlocal conductance does not take substantial positive values.
Nevertheless, in the unshaded region, 
the positive nonlocal conductance persists
as long as the altermagnet retains the FS configuration shown in Fig.~\ref{figure1}(c).
Similarly, Fig.~\ref{figure6}(c) shows the zero-bias differential conductance $G_1$ as a function of the on-site energy for the $p_x$ and $p_y$ orbitals ($\varepsilon_2$).
In the shaded region satisfying $\varepsilon_2>-\mu-4t_2^2/(m_2-\mu)$, 
the configuration of the altermagnet FSs changes qualitatively; 
for instance, 
additional altermagnetic FSs enclosing the $\Gamma$ point emerge, 
as illustrated for $\varepsilon_2=t_1$ in Fig.~\ref{figure6}(d).
In this region, as shown in Fig.~\ref{figure6}(c), 
the positive nonlocal conductance is suppressed.
In contrast, outside the shaded region, 
the altermagnet retains the characteristic spin-split FS geometry as shown in Fig.~\ref{figure1}(c),
thereby preserving the positive signal.
Overall, these results demonstrate that the positive nonlocal conductance persists over a wide parameter range.


\subsection{Dependence on lead positions}

\begin{figure}[t]
\begin{center}
\includegraphics[width=0.5\textwidth]{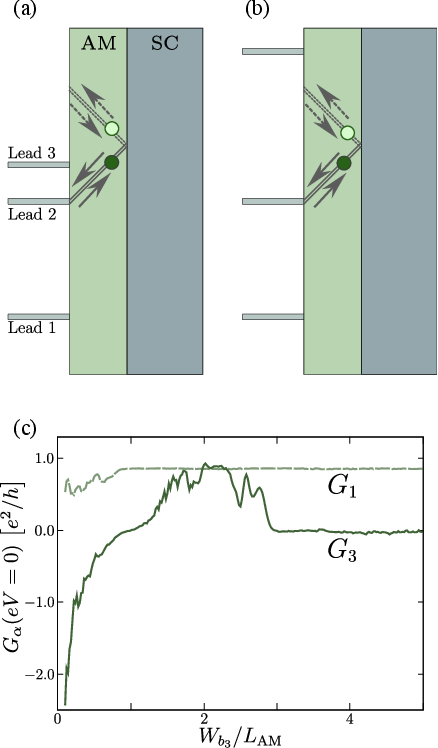}
\caption{(a), (b) Schematic illustrations of scattering processes when the distance between the second and third leads deviates from $W_{b_3}/L_{\textrm{AM}}\sim 2$.
(c) Zero-bias differential conductance as a function of the distance between the second and third leads.}
\label{figure7}
\end{center}
\end{figure}

In this section, we investigate how the placement of the normal leads affects the nonlocal conductance.
As suggested by Figs.~\ref{figure3}(e)--(h), spin-up (spin-down) holes generated via SAR
can be efficiently extracted into the normal leads when $W_{b_2}/L_{\textrm{AM}}\sim 2$ ($W_{b_3}/L_{\textrm{AM}}\sim 2$).
For example, when the third lead is either too close to or too far from the second lead,
the reflected holes fail to enter the third lead, as illustrated in Figs.~\ref{figure7}(a)~and~\ref{figure7}(b), respectively.
Figure~\ref{figure7}(c) shows the numerically computed nonlocal conductance at zero bias voltage as a function of the distance between the second and third leads (i.e., $W_{b_3}$),
where the position of first lead is fixed to satisfy $W_{b_2}/L_{\textrm{AM}}\sim 2$ for comparison.
Specifically, we choose the parameters $W_\textrm{N}=5a', W_\textrm{AM}=600a',W_\textrm{S}=599a', L_{\textrm{AM}} = 50a', W_{b_1}=200a'$, and $W_{b_2}=100a'$.
At the interface, we assume orbital-independent hopping integrals, $\bar{t}_{J_\textrm{N(S)},\alpha}=0.7t_1$ and ${\delta t_{J_\textrm{N(S)},\alpha}}=0.5t_1$.
The nonlocal conductance $G_1$ is positive and almost independent of $W_{b_3}$.
This result demonstrates that the positive nonlocal conductance between the two optimally spaced leads
is robust against modifications of the interference pattern induced by the positions of the other leads.
In contrast, $G_3$ exhibits a significant dependence on $W_{b_3}$ and takes substantial positive values only around $W_{b_3}/L_\textrm{AM}\sim 2$.
When the distance between the third and second leads becomes small,
electrons undergoing retroreflective normal reflection at the altermagnet--superconductor interface can leak into the third lead.
This leakage likely accounts for the negative sign of $G_3$ and the slight reduction in $G_1$ for smaller $W_{b_3}$.
The conductance $G_{\alpha}$ also exhibits rapid, irregular oscillations, which may arise from quantum interference effects sensitive to the system configuration~\cite{stone_90}.
In practical experiments, more than three normal leads can be attached.
Positive nonlocal conductance appears only when the distance between
the biased lead (through which electrons are injected) and the probed lead (where the nonlocal current is measured),
$\delta W$, satisfies $\delta W/L_{\textrm{AM}} \sim 2$.
By modifying the bias-voltage configuration, one can interchange the roles of the biased and probed leads without altering the device geometry.
Consequently, both the presence and absence of SAR channels can be tested within a single multiterminal device, corresponding to the primary and control experiments, respectively.
In a realistic setup, residual crossed Andreev reflection may also contribute to the nonlocal conductance.
However, its contribution is expected to decrease monotonically as the lead separation $\delta W$ increases.
In contrast, as shown in Fig.~\ref{figure7}, the interlead Andreev scattering is maximized around $\delta W/L_{\mathrm{AM}} = 2$.
Therefore, the markedly different dependence on $\delta W$ provides a useful criterion for distinguishing SAR from crossed Andreev reflection.

\section{Discussion and Conclusion}

A recent neutron diffraction experiment reports the absence of altermagnetism in bulk KV$_2$Se$_2$O~\cite{shiliangli_2025},
suggesting that surface altermagnetism may instead be realized in this compound~\cite{moo_2026}.
As demonstrated in Fig.~\ref{figure5}(a), we observe substantial positive nonlocal conductance
when the normal and superconductor leads are coupled only to the $d_{xz}$, $d_{yz}$, and $p_z$ orbitals that extend along the out-of-plane direction.
Thus, we expect that SAR can be realized even with surface altermagnetism by attaching the normal and superconducting leads to the top surface of KV$_2$Se$_2$O.
However, a detailed validation of this idea, which requires explicit three-dimensional calculations, will be addressed in future work.
In this paper, we neglect spin-flip scattering induced by weak spin--orbit coupling.
Previous studies based on a minimal single-orbital model suggest that SAR is not immediately suppressed by weak spin--orbit coupling\cite{Nagae_2025}. 
A quantitative analysis of spin--orbit coupling effects is left for future work.
In our calculations, the altermagnet is assumed to be oriented at an angle of $45^\circ$ with respect to the normal metals and the superconductor.
Nevertheless, as illustrated in Fig.~\ref{figure3}, the occurrence of SAR for a $45^\circ$ interface orientation is not dictated by symmetry,
but rather results from favorable matching between the quasiparticle group velocities and the positions of the leads.
We therefore expect that positive nonlocal conductance remains observable even in the presence of a small angular misalignment.
A systematic investigation of the dependence on the interface orientation, as well as the surface termination, is an important direction for future work.
In realistic altermagnets, magnetic or twin domains may be present.
Such domain walls can induce retroreflective normal scattering, thereby substantially reducing the probability that an incident electron reaches the altermagnet--superconductor interface.
As a result, magnetic or twin domains may significantly suppress the positive nonlocal conductance arising from dominant SAR.
These considerations suggest that high-quality single-domain altermagnetic samples are desirable for the robust observation of SAR.
A microscopic investigation of the effects of domain walls, however, is beyond the scope of the present work and remains an important subject for future study.
Although positive nonlocal conductance provides strong evidence for dominant SAR,
spin-resolved transport measurements using spin-polarized current injection~\cite{Jedema_2001,Ando_2014,Tian_2015}
would offer a more direct test of the predicted scattering processes illustrated in Fig.~\ref{figure3}.
Furthermore, although experimentally challenging, spatial current imaging of the altermagnet could directly visualize the current path associated with SAR
~\cite{Roth_1989,Xing_1994,Knauss_2001,Schrag_2003,Nowack_2013,Shibata_2015,Tetienne_2017,Zhang_2020}.

In summary, we study the transport properties of a junction between a V$_2$O-based altermagnet and a conventional superconductor,
where the altermagnet is described by a microscopically motivated seven-basis-state model with six-orbital characters
that explicitly incorporates both magnetic V sites and O sites.
By calculating the nonlocal differential conductance under various boundary conditions, we demonstrate the robust emergence of spin-polarized SAR at the junction interface.
The proposed setup allows us to probe both the presence and absence of SAR channels within a single device, providing an efficient platform for primary and control experiments.
We anticipate that this work will motivate future experimental efforts toward realizing spin-resolved Cooper-pair splitting in V$_2$O-based altermagnets.

\acknowledgements
We thank Y. Maeno for the insightful discussions. 
Y.N. is supported by JST SPRING (Grant No. JPMJSP2125) and thanks the THERS Make New Standards Program for the Next Generation Researchers and a Grant-in-Aid for JSPS Fellows (JSPS KAKENHI Grant No. JP26KJ1305).
S.I. is supported by a Grant-in-Aid for Early-Career Scientists (JSPS KAKENHI Grant No. JP24K17010).
A.P.S. is funded by the Deutsche Forschungsgemeinschaft (DFG, German Research Foundation) -- TRR 360 -- 492547816. 

\section*{DATA AVAILABILITY}
The data that support the findings of this paper are openly available\cite{Data_availability}.

\appendix
\section{BdG Hamiltonian for \\ the multiterminal setup}\label{sec:appendixa}
In this appendix, we present the explicit form of the Hamiltonian used to compute the differential conductance.
The Hamiltonian in the altermagnet region shown in Fig.~\ref{figure2}(a) is given by
\begin{widetext}
	\begin{align}
		\begin{split}
			&H_\textrm{AM} = \sum_{\sigma=\uparrow,\downarrow}H_{\textrm{AM},\sigma}, \\
			&H_{\textrm{AM},\sigma} = \frac{1}{2}\sum_{j=0}^{N_L-1}\left\{\qty(
				\qty[\vb*{\Psi}_{j+\frac{1}{2},\sigma,B}^\dagger, \vb*{\Psi}_{j+\frac{1}{2},\bar{\sigma},B}^\textrm{T}]
				\mqty[\check{T}_{BA} & \\ & -\check{T}_{BA}^*]\mqty[\vb*{\Psi}_{j,\sigma,A}\\ \vb*{\Psi}_{j,\bar{\sigma},A}^*]+\mathrm{H.c.}) \right.\\
				&\quad\quad +\qty(\qty[\vb*{\Psi}_{j+1,\sigma,A}^\dagger,\vb*{\Psi}_{j+1,\bar{\sigma},A}^\textrm{T}]\mqty[\check{T}_{AB} & \\ & -\check{T}_{AB}^*]\mqty[\vb*{\Psi}_{j+\frac{1}{2},\sigma,B} \\ \vb*{\Psi}_{j+\frac{1}{2},\bar{\sigma},B}^*]+\mathrm{H.c.}) \\
				&\quad\quad\left.+\qty[\vb*{\Psi}_{j+\frac{1}{2},\sigma,B}^\dagger, \vb*{\Psi}_{j+\frac{1}{2},\bar{\sigma},B}^\textrm{T}]\mqty[\check{\mathcal{E}}_{B,\sigma} & \\ & -\check{\mathcal{E}}_{B,\bar{\sigma}}^*]\mqty[\vb*{\Psi}_{j+\frac{1}{2},\sigma,B} \\ \vb*{\Psi}_{j+\frac{1}{2},\bar{\sigma},B}^*]\right\}
				+\sum_{j=0}^{N_L}\qty[\vb*{\Psi}_{j,\sigma,A}^\dagger,\vb*{\Psi}_{j,\sigma,A}^\textrm{T}]\mqty[\check{\mathcal{E}}_{A,\sigma} & \\ & -\check{\mathcal{E}}_{A,\bar{\sigma}}^*]\mqty[\vb*{\Psi}_{j,\sigma,A} \\ \vb*{\Psi}_{j,\bar{\sigma},A}^*],\\
			&\Psi_{j,\sigma,A(B)}^\dagger = \qty[\vb*{\Psi}_{\sigma,A(B),p_z,j}^\dagger, \vb*{\Psi}_{\sigma,A(B),V,j+\frac{1}{4}}^\dagger,\vb*{\Psi}_{\sigma,A(B),p_x,j}^\dagger,\vb*{\Psi}_{\sigma,A(B),p_y,j}^\dagger,\vb*{\Psi}_{j+\frac{1}{4},\sigma,A(B),d_{xy}}^\dagger], \\
			&\vb*{\Psi}_{j,\sigma,A(B),p_\eta}^\dagger = \qty[c_{j,0(\frac{1}{2})\sigma,p_\eta}^\dagger, \ldots, c_{j,N_W-1 \ (N_W-\frac{3}{2})\sigma,p_\eta}^\dagger],\\
			&\vb*{\Psi}_{j,\sigma,A(B),V}^\dagger = \qty[c_{j,\frac{1}{4}\sigma,d_{yz(xz)}}^\dagger,c_{j,\frac{3}{4},\sigma,d_{xz(yz)}}^\dagger, \ldots, c_{j,N_W-\frac{7}{4},\sigma,d_{yz(xz)}}^\dagger, c_{j,N_W-\frac{5}{4},\sigma,d_{xz(yz)}}^\dagger],\\
			&\vb*{\Psi}_{j,\sigma,A(B),d_{xy}}^\dagger = \qty[c_{j,\frac{1}{4},\sigma,d_{xy}V_{\uparrow(\downarrow)}}^\dagger, c_{j,\frac{3}{4},\sigma,d_{xy}V_{\downarrow(\uparrow)}}^\dagger,\ldots,c_{j,N_W-\frac{7}{4},\sigma,d_{xy}V_{\uparrow(\downarrow)}}^\dagger,c_{j,N_W-\frac{5}{4},\sigma,d_{xy}V_{\downarrow(\uparrow)}}^\dagger],\\
			&\check{\mathcal{E}}_{\sigma,A}
			= \mqty[\hat{\mathcal{E}}_{p_z,A} & \hat{T}_{V\textrm{-}p_z,A}^\dagger & 0 & 0 & 0\\
							\hat{T}_{V\textrm{-}p_z,A} & \hat{\mathcal{E}}_{\sigma,V,A} & 0 & 0 & 0 \\
							0 & 0 & \hat{\mathcal{E}}_{p_x,A} & 0 & \hat{T}_{d_{xy}\textrm{-}p_x,A}^\dagger \\
							0 & 0 & 0 & \hat{\mathcal{E}}_{p_y,A} & \hat{T}_{d_{xy}\textrm{-}p_y,A}^\dagger \\
							0 & 0 & \hat{T}_{d_{xy}\textrm{-}p_x,A} & \hat{T}_{d_{xy}\textrm{-}p_y,A} & \hat{\mathcal{E}}_{\sigma,d_{xy},A}],\\
			&\check{\mathcal{E}}_{\sigma,B}
			= \mqty[\hat{\mathcal{E}}_{p_z,B} & \hat{T}_{V\textrm{-}p_z,B}^\dagger & 0 & 0 & 0\\
							\hat{T}_{V\textrm{-}p_z,B} & \hat{\mathcal{E}}_{\sigma,V,B} & 0 & 0 & 0 \\
							0 & 0 & \hat{\mathcal{E}}_{p_x,B} & 0 & \hat{T}_{d_{xy}\textrm{-}p_x,B}^\dagger \\
							0 & 0 & 0 & \hat{\mathcal{E}}_{p_y,B} & \hat{T}_{d_{xy}\textrm{-}p_y,B}^\dagger \\
							0 & 0 & \hat{T}_{d_{xy}\textrm{-}p_x,B} & \hat{T}_{d_{xy}\textrm{-}p_y,B} & \hat{\mathcal{E}}_{\sigma,d_{xy},B}],\\
			&\check{T}_{BA}
			= \mqty[0 & \hat{T}_{p_z\textrm{-}V,BA} & 0 & 0 & 0 \\
							0 & 0												 & 0 & 0 & 0 \\
							0 & 0												 & 0 & 0 & \hat{T}_{p_x\textrm{-}d_{xy},BA} \\
							0 & 0												 & 0 & 0 & \hat{T}_{p_y\textrm{-}d_{xy},BA} \\
							0 & 0												 & 0 & 0 & \hat{T}_{d_{xy}\textrm{-}d_{xy},BA}], 
			\check{T}_{AB}
			= \mqty[0 & \hat{T}_{p_z\textrm{-}V,AB} & 0 & 0 & 0 \\
							0 & 0												 & 0 & 0 & 0 \\
							0 & 0												 & 0 & 0 & \hat{T}_{p_x\textrm{-}d_{xy},AB} \\
							0 & 0												 & 0 & 0 & \hat{T}_{p_y\textrm{-}d_{xy},AB} \\
							0 & 0												 & 0 & 0 & \hat{T}_{d_{xy}\textrm{-}d_{xy},AB}], \\
			&(\hat{\mathcal{E}}_{p_z,A(B)})_{k,l} = (-\varepsilon_1-\mu)\delta_{k,l},\ (\hat{\mathcal{E}}_{p_{x(y)},A(B)})_{k,l} = (-\varepsilon_2-\mu)\delta_{k,l}, \\
			&(\hat{\mathcal{E}}_{\sigma,V,A(B)})_{k,l} = (-s_ks_{\alpha}s_{\sigma}m_1-\mu)\delta_{k,l},\
			(\hat{\mathcal{E}}_{\sigma,d_{xy},A(B)})_{k,l} = (-s_ks_{\alpha}s_{\sigma}m_2-\mu)\delta_{k,l} - t_3\delta_{|k-l|,1},\\
			&(\hat{T}_{V\textrm{-}p_z,A})_{k,l} = t_1(\delta_{k-2l,-1}+\delta_{k-2l,-2}) = (-\hat{T}_{p_z\textrm{-}V,AB})_{l,k},\
			(\hat{T}_{d_{xy}\textrm{-}p_x,A})_{k,l} = t_2\delta_{k-2l,-1} = (-\hat{T}_{p_x\textrm{-}d_{xy},AB})_{l,k},\\
			&(\hat{T}_{p_y\textrm{-}d_{xy},A})_{k,l} = t_2\delta_{k-2l,-2} = (-\hat{T}_{p_y\textrm{-}d_{xy},AB})_{l,k},\
			(\hat{T}_{V\textrm{-}p_z,B})_{k,l} = t_1(\delta_{k-2l,-1}+\delta_{k-2l,0}) = (-\hat{T}_{p_z\textrm{-}V,BA})_{l,k}, \\
			&(\hat{T}_{d_{xy}\textrm{-}p_x,B})_{k,l} = t_2\delta_{k-2l,-1} = (-\hat{T}_{p_x\textrm{-}d_{xy},BA})_{l,k},\
			(\hat{T}_{d_{xy}\textrm{-}p_y,B})_{k,l} = t_2\delta_{k-2l,0} = (-\hat{T}_{p_y\textrm{-}d_{xy},BA})_{l,k},\\
			&(\hat{T}_{d_{xy}\textrm{-}d_{xy},BA(AB)})_{k,l} = -t_3\delta_{k,l},
		\end{split}
	\end{align}
\end{widetext}
where, $c_{j,m,\sigma,\alpha}^\dagger$ ($c_{j,m,\sigma,\alpha}$) denotes the creation (annihilation) operator for an electron with spin $\sigma$
and orbital $\alpha$ ($= p_x, p_y, p_z, d_{xz}, d_{yz}, d_{xy}\mathrm{V}\uparrow, d_{xy}\mathrm{V}\downarrow$) at position $\vb*{r} = j \vb*{a}_{x}' + m \vb*{a}_{y}'$.
We define the primitive lattice vectors as $\vb*{a}_{x}' = \vb*{a}_x + \vb*{a}_y$ and $\vb*{a}_{y}' = -\vb*{a}_x + \vb*{a}_y$.
The system size is specified by integers $N_L$ and $N_W$, such that $L_{\textrm{AM}} = a' N_L$ and $W_{\textrm{AM}} = a' N_W$.
We introduce $s_k = +1,(-1)$ for odd (even) $k$, $s_{\alpha} = +1,(-1)$ for $\alpha = A,(B)$, and $s_{\sigma} = +1,(-1)$ for $\sigma = \uparrow,(\downarrow)$.
$\hat{\mathcal{E}}_{p_\eta,A(B)}$ is an $N_W \times N_W$ ($[N_W-1] \times [N_W-1]$) matrix for $\eta = x,y,z$.
$\hat{\mathcal{E}}_{V(d_{xy}),A(B),\sigma}$ is a $2(N_W-1) \times 2(N_W-1)$ matrix.
$\hat{T}_{V\textrm{-}p_z(d_{xy}\textrm{-}p_\eta),A}$ is a $2(N_W-1) \times N_W$ matrix for $\eta = x,y$.
$\hat{T}_{V\textrm{-}p_z(d_{xy}\textrm{-}p_\eta),B}$ is a $2(N_W-1) \times (N_W-1)$ matrix for $\eta = x,z$.
$\hat{T}_{p_z\textrm{-}V(p_\eta\textrm{-}d_{xy}),BA}$ is an $(N_W-1) \times 2(N_W-1)$ matrix for $\eta = x,y$.
$\hat{T}_{d_{xy}\textrm{-}d_{xy},BA(AB)}$ is a $2(N_W-1) \times 2(N_W-1)$ matrix.
$\hat{T}_{p_z\textrm{-}V(p_\eta\textrm{-}d_{xy}),AB}$ is an $N_W \times 2(N_W-1)$ matrix.
The Hamiltonian for the normal metals is given by,
\begin{widetext}
	\begin{align}
		\begin{split}
			&H_{\textrm{N}} = \frac{1}{2}\sum_{\sigma}\sum_{j}\qty{\qty(\qty[\vb*{\Psi}_{j+1,\sigma,\textrm{N}}^\dagger,\vb*{\Psi}_{j+1,\bar{\sigma},\textrm{N}}^\textrm{T}] \mqty[\check{T}_{\textrm{N}} & \\ & -\check{T}_{\textrm{N}}^*]\mqty[\vb*{\Psi}_{j,\sigma,\textrm{N}} \\ \vb*{\Psi}_{j,\bar{\sigma},\textrm{N}}^*] +\mathrm{H.c.}) + \qty[\vb*{\Psi}_{j,\sigma,\textrm{N}}^\dagger,\vb*{\Psi}_{j,\bar{\sigma},\textrm{N}}^\textrm{T}] \mqty[\check{\mathcal{E}}_{\textrm{N}} & \\ & -\check{\mathcal{E}}_{\textrm{N}}^*] \mqty[\vb*{\Psi}_{j,\sigma,\textrm{N}} \\ \vb*{\Psi}_{j,\bar{\sigma},\textrm{N}}^*]}, \\
			&\vb*{\Psi}_{j,\sigma,\textrm{N}}^\dagger = \qty[\vb*{\Psi}_{j,\sigma,\textrm{N}_1}^\dagger, \vb*{\Psi}_{j,\sigma,\textrm{N}_2,}^\dagger, \vb*{\Psi}_{j,\sigma,\textrm{N}_3}^\dagger],
			\vb*{\Psi}_{j,\sigma,\textrm{N}_n}^\dagger = \qty[b_{j,1,\sigma,n}^\dagger, \ldots,b_{j,N_\textrm{N},\sigma,n}^\dagger],
			\check{\mathcal{E}}_{\textrm{N}} = \mqty[\hat{\mathcal{E}}_{\textrm{N}} & & \\ & \hat{\mathcal{E}}_{\textrm{N}} & \\ & & \hat{\mathcal{E}}_{\textrm{N}}], 
			\check{T}_{\textrm{N}} = \mqty[\hat{T}_{\textrm{N}} & & \\ & \hat{T}_{\textrm{N}} & \\ & & \hat{T}_{\textrm{N}}], \\
			&(\hat{\mathcal{E}}_{\textrm{N}})_{k,l} = -\mu_\textrm{N}\delta_{k,l} - t_\textrm{N}\delta_{|k-l|,1}, \ (\hat{T}_{\textrm{N}})_{k,l} = -t_{\textrm{N}}\delta_{k,l},
		\end{split}
	\end{align}
\end{widetext}
where $b_{j,m,\sigma,n}^\dagger(b_{j,m,\sigma,n})$ represents the creation (annihilation) operator of an electron
with spin $\sigma$ at $\vb*{r} = j\vb*{a}_x'+m\vb*{a}_y'$ in the $n$-th lead,
$\mu_\textrm{N}$ is the chemical potential and $t_\textrm{N}$ denotes the hopping integral.
$\hat{\mathcal{E}}_{\textrm{N}}$ and $\hat{T}_{\textrm{N}}$ are $N_\textrm{N} \times N_\textrm{N}$ matrices,
where we define  $N_\textrm{N} \equiv W_\textrm{N}/a_\textrm{N}$. 
The Hamiltonian for the superconductor is given by,
\begin{widetext}
	\begin{align}
		\begin{split}
			&H_{\textrm{S}} = \frac{1}{2}\sum_{\sigma}\sum_{j}\qty[\qty(
				\mqty[\vb*{\Psi}_{j+1,\sigma,\textrm{S}}^\dagger, \vb*{\Psi}_{j+1,\bar{\sigma},\textrm{S}}^{\mathrm{T}}]\mqty[\hat{T}_{\textrm{S}} & \\ & -\hat{T}_{\textrm{S}}^*]\mqty[\vb*{\Psi}_{j,\sigma,\textrm{S}} \\ \vb*{\Psi}_{j,\bar{\sigma},\textrm{S}}^*] + \mathrm{H.c.}) 
				+ \mqty[\vb*{\Psi}_{j,\sigma,\textrm{S}}^\dagger, \vb*{\Psi}_{j,\bar{\sigma},\textrm{S}}^{\mathrm{T}}]
				\mqty[\hat{\mathcal{E}}_{\textrm{S}} & s_\sigma \hat{\Delta} \\ s_\sigma \hat{\Delta} & -\hat{\mathcal{E}}_{\textrm{S}}^*]
				\mqty[\vb*{\Psi}_{j,\sigma,\textrm{S}} \\ \vb*{\Psi}_{j,\bar{\sigma},\textrm{S}}^*]], \\
			&\vb*{\Psi}_{j,\sigma,\textrm{S}}^\dagger = \qty[b_{j,1,\sigma,\textrm{S}}^\dagger, \ldots, b_{j,N_{\textrm{S}},\sigma,\textrm{S}}^\dagger], 
			(\hat{\mathcal{E}}_{\textrm{S}})_{k,l} = -\mu_\textrm{S}\delta_{k,l} - t_\textrm{S}\delta_{|k-l|,1}, 
			\hat{\Delta}_{k,l} = \Delta \delta_{k,l}, 
			(\hat{T}_{\textrm{S}})_{k,l} = -t_\textrm{S}\delta_{k,l},
		\end{split}
	\end{align}
\end{widetext}
where $b_{j,m,\sigma,\textrm{S}}^\dagger(b_{j,m,\sigma,\textrm{S}})$ represents the creation (annihilation) operator of an electron
with spin $\sigma$ at $\vb*{r}=(j/2)\vb*{a}_x'+(m/2)\vb*{a}_y'$ in the superconductor,
$\mu_\textrm{S}$ is the chemical potential, $t_\textrm{S}$ denotes the hopping integral and $\Delta$ is the pair potential.
$s_\sigma=+1(-1)$ for $\sigma = \uparrow(\downarrow)$.
$\hat{\mathcal{E}}_{\textrm{S}}$, $\hat{T}_{\textrm{S}}$ and $\hat{\Delta}$ are $N_\textrm{S} \times N_\textrm{S}$ matrices, where we define  $N_\textrm{S} \equiv W_\textrm{S}/a_\textrm{S}$. 
The Hamiltonian at the interfaces between the normal metals and the altermagnet is given by,
\begin{widetext}
	\begin{align}
		\begin{split}
			&H_{\textrm{AM-N}} = \frac{1}{2}\sum_{\sigma=\uparrow,\downarrow}\qty(\qty[\vb*{\Psi}_{0,\sigma,A}^\dagger,\vb*{\Psi}_{0,\bar{\sigma},A}^\textrm{T}]\mqty[\check{T}_{\textrm{AM-N}} & \\ & -\check{T}_{\textrm{AM-N}}^*]\mqty[\vb*{\Psi}_{-1,\sigma,\textrm{N}} \\ \vb*{\Psi}_{-1,\bar{\sigma},\textrm{N}}^*] + \mathrm{H.c.}), \\
			&\check{T}_{\textrm{AM-N}} = \mqty[\hat{T}_{\textrm{AM-N}_1,p_z} & \hat{T}_{\textrm{AM-N}_2,p_z} & \hat{T}_{\textrm{AM-N}_3,p_z}\\
																				 \hat{T}_{\textrm{AM-N}_1,V} & \hat{T}_{\textrm{AM-N}_2,V} & \hat{T}_{\textrm{AM-N}_3,V}\\
																				 \hat{T}_{\textrm{AM-N}_1,p_x} & \hat{T}_{\textrm{AM-N}_2,p_x} & \hat{T}_{\textrm{AM-N}_3,p_x}\\
																				 \hat{T}_{\textrm{AM-N}_1,p_y} & \hat{T}_{\textrm{AM-N}_2,p_y} & \hat{T}_{\textrm{AM-N}_3,p_y}\\
																				 \hat{T}_{\textrm{AM-N}_1,d_{xy}} & \hat{T}_{\textrm{AM-N}_2,d_{xy}} & \hat{T}_{\textrm{AM-N}_3,d_{xy}}], \\
			&(\hat{T}_{\textrm{AM-N}_1,p_\eta})_{k,l} = -t_{J_\textrm{N},p_\eta;\vb*{r},\vb*{r}'}\delta_{k-l,N_{b_1}},\\
			&(\hat{T}_{\textrm{AM-N}_2,p_\eta})_{k,l} = -t_{J_\textrm{N},p_\eta;\vb*{r},\vb*{r}'}\delta_{k-l,N_{b_1}+N_{b_2}+N_\textrm{N}},\\
			&(\hat{T}_{\textrm{AM-N}_3,p_\eta})_{k,l} = -t_{J_\textrm{N},p_\eta;\vb*{r},\vb*{r}'}\delta_{k-l,N_{b_1}+N_{b_2}+N_{b_3}+2N_\textrm{N}},\\
			&(\hat{T}_{\textrm{AM-N}_1,V(d_{xy})})_{k,l} = -t_{J_\textrm{N},V(d_{xy});\vb*{r},\vb*{r}'}\qty(\delta_{k-2l,2(N_{b_1}-1)}+\delta_{k-2l,2N_{b_1}-1}),\\
			&(\hat{T}_{\textrm{AM-N}_2,V(d_{xy})})_{k,l} = -t_{J_\textrm{N},V(d_{xy});\vb*{r},\vb*{r}'}\qty(\delta_{k-2l,2(N_{b_1}+N_{b_2}+N_\textrm{N}-1)}+\delta_{k-2l,2(N_{b_1}+N_{b_2}+N_\textrm{N})-1}),\\
			&(\hat{T}_{\textrm{AM-N}_3,V(d_{xy})})_{k,l} = -t_{J_\textrm{N},V(d_{xy});\vb*{r},\vb*{r}'}\qty(\delta_{k-2l,2(N_{b_1}+N_{b_2}+N_{b_3}+2N_\textrm{N}-1)}+\delta_{k-2l,2(N_{b_1}+N_{b_2}+N_{b_3}+2N_\textrm{N})-1}),
		\end{split}
	\end{align}
\end{widetext}
where we define $N_{b_1}$, $N_{b_2}$ and $N_{b_3}$ as $W_{b_1}-W_\textrm{N}/2=N_{b_1}a'$, $W_{b_2}-W_\textrm{N}=N_{b_2}a'$, and $W_{b_3}-W_\textrm{N}=N_{b_3}a'$.
$\hat{T}_{\textrm{AM-N}_n,p_\eta}$ is an $N_W\times N_\textrm{N}$ matrix for $\eta=x,y,z$.
$\hat{T}_{\textrm{AM-N}_n,V(d_{xy})}$ is a $2(N_W-1)\times N_\textrm{N}$ matrix.
The Hamiltonian at the interface between the altermagnet and the superconductor is given by,
\begin{widetext}
	\begin{align}
		\begin{split}
			&H_{\textrm{S-AM}} = \frac{1}{2}\sum_{\sigma=\uparrow,\downarrow}\qty(\qty[\vb*{\Psi}_{N_L+1,\sigma,\textrm{S}}^\dagger,\vb*{\Psi}_{N_L+1,\bar{\sigma},\textrm{S}}^\textrm{T}] \mqty[\check{T}_{\textrm{S-AM}} & \\ & -\check{T}_{\textrm{S-AM}}^*]\mqty[\vb*{\Psi}_{N_L,\sigma,A} \\ \vb*{\Psi}_{N_L,\bar{\sigma},A}^*] + \mathrm{H.c.}), \\
			&\check{T}_{\textrm{S-AM}} = \mqty[\hat{T}_{\textrm{S-AM},p_z} & \hat{T}_{\textrm{S-AM},V}& \hat{T}_{\textrm{S-AM},p_x}& \hat{T}_{\textrm{S-AM},p_y}& \hat{T}_{\textrm{S-AM},d_{xy}}],\\
			&(\hat{T}_{\textrm{S-AM},p_\eta})_{k,l} = -t_{J_\textrm{S},p_\eta;\vb*{r},\vb*{r}'}\qty(\delta_{k-2l+1,0}+\delta_{k-2l+2,0}),\\
			&(\hat{T}_{\textrm{S-AM},V(d_{xy})})_{k,l} = -t_{J_\textrm{S},V(d_{xy});\vb*{r},\vb*{r}'}\delta_{k-l,0},
		\end{split}
	\end{align}
\end{widetext}
where $\hat{T}_{\textrm{S-AM},p_\eta}$ is a $2(N_\textrm{S}-1)\times N_W$ matrix for $\eta = x,y,z$.
$\hat{T}_{\textrm{S-AM},V(d_{xy})}$ is a $2(N_\textrm{S}-1)\times 2(N_W-1)$ matrix.

\end{document}